\documentclass[a4paper,11pt,fleqn]{article}
\usepackage{authblk}
\usepackage{datetime}

\usepackage{pstricks}
\usepackage{colortbl}

\usepackage{pst-node}
\usepackage{pst-grad}
\usepackage{pst-coil}
\usepackage{pst-text}
\usepackage{pst-rel-points}
\usepackage{amssymb}
\usepackage{amsmath}
\usepackage{mathptm}
\usepackage{calc}
\usepackage{etex}
\usepackage{ifthen}
\usepackage{multirow}
\usepackage{rotating}
\usepackage{longtable}
\usepackage[normalem]{ulem}
\usepackage{stmaryrd}
\usepackage{xspace}

\newcommand{\End}[1]{\text{{\sf\em E}$_{#1}$}\xspace}
\newcommand{\Start}[1]{\text{{\sf\em S}$_{#1}$}\xspace}

\newcommand{\pred}{\text {\sf\em{pred}\/}}
\newcommand{\Succ}{\text{\sf\em {succ}\/}}

\newcommand{\var}{\text{\sf\em V\/}\xspace}
\newcommand{\expr}{\text{\sf\em E\/}\xspace}
\newcommand{\const}{\text{\sf\em C\/}\xspace}
\newcommand{\pointer}{\text{\sf\em P\/}\xspace}

\newdateformat{titledate}{\THEDAY\ \monthname[\THEMONTH], \THEYEAR}
\newdateformat{mydate}{\monthname[\THEMONTH] \THEYEAR}

\psset{unit=1mm}

\newcommand{\lin}{\text{{\sf\em Lin\/}}\xspace}
\newcommand{\ain}{\text{{\sf\em Ain\/}}\xspace}

\newcommand{\lout}{\text{{\sf\em Lout\/}}\xspace}
\newcommand{\aout}{\text{{\sf\em Aout\/}}\xspace}

\newcommand{\must}{\text{\sf\em Must\/}\xspace}

\newcommand{\Def}{\text{{\sf\em Def}}\xspace}
\newcommand{\Pointee}{\text{{\sf\em Pointee}}\xspace}
\newcommand{\Kill}{\text{{\sf\em Kill}}\xspace}
\newcommand{\Ref}{\text{{\sf\em Ref}}\xspace} 

\newcommand{\lrestrict}[2]{\text{$\protect#1{\mid}_{\protect#2}$}}

\protect{\newlength{\BRACEL}}
\protect{\newlength{\TAL}} 
\newlength{\codeLineLength}

\newcommand{\eval}[1]{\text{$\llbracket{#1}\rrbracket$}}
\newcommand{\app}{\text{\sf\em approx\/}\xspace}
\newcommand{\isunion}{\text{\sf\em union\/}}

\begin{document}

\date{}

\title{\textbf{Generalizing the Liveness Based Points-to Analysis}}
\author{{\Large Uday P. Khedker} and {\Large Vini Kanvar}\thanks{Vini has been partially supported by the TCS Research Fellowship.}}
\affil{Department of Computer Science and Engineering \\ Indian Institute of Technology Bombay \\ 
Email: \tt\{{uday,vini\}@cse.iitb.ac.in}}
\titledate
\maketitle
\thispagestyle{empty}

\begin{abstract}
The original liveness based flow and context sensitive points-to
analysis (LFCPA) is restricted to scalar pointer variables and scalar
pointees on stack and static memory. In this paper, we extend it to
support heap memory and pointer expressions involving structures, unions,
arrays, and pointer arithmetic. The key idea behind these extensions
involves constructing bounded names for locations in terms of compile
time constants (names and fixed offsets), and introducing sound
approximations when it is not possible to do so. We achieve this by
defining a grammar for pointer expressions, suitable memory models
and location naming conventions, and some key evaluations of pointer
expressions that compute the named locations. These extensions preserve
the spirit of the original LFCPA which is evidenced by the fact that
although the lattices and the location extractor functions change, the 
data flow equations remain unchanged.
\end{abstract}

\section{Introduction}
The liveness based flow and context sensitive pointer analysis
(LFCPA)~\cite{lfcpa} was proposed as a way of containing the
combinatorial explosion of points-to information. Intuitively, it
enabled garbage collection over points-to information---the points-to
pairs that are guaranteed to be unusable 
are removed from the points-to relations. 
This is achieved by restricting the propagation of
points-to information to the live ranges of pointers which 
eliminates the points-to information of pointers that are dead. 

For simplicity of exposition, LFCPA was formulated for stack/static
memory and scalar pointers although the implementation handled
heap, structures, arrays, and pointer arithmetic conservatively but
in somewhat adhoc manner. This paper is an attempt to formalize
the extensions of LFCPA to heap memory, structures, arrays, 
pointer arithmetic, and unions. 
The key idea behind these extensions
involves constructing bounded names for locations in terms of compile
time constants (names and fixed offsets), and introducing sound
approximations when it is not possible to do so. 
We achieve this by
defining a grammar for pointer expressions, suitable memory models
and location naming conventions, and some key evaluations of pointer
expressions that compute the named locations.

Our naming conventions are based on using variables names, allocation site names, and sequences
involving field names or constant offsets. All these are used to create bounded names
of the memory locations of interest. Our approximations include
using allocation site names, using a collection
of named locations, and dropping field names to approximate unions field insensitively.
The actual naming conventions and approximations depends on the choice of memory model.
The evaluations of pointer expressions
include computing their l- and r-values and computing the set of pointers dereferenced to 
reach the l- or the r-value. 


The focus of this paper is on declarative formulations of the
extensions rather than efficient algorithms for computing the points-to
information. Further, this paper should be seen as a follow up work of LFCPA rather than
an independently understandable description. We provide only a brief
summary of the original LFCPA in Section~\ref{sec:original.lfcpa};
please see~\cite{lfcpa} for more details of the original formulation
of LFCPA. Section~\ref{sec:lfcpa.heap} extends it to support the
heap memory and structures (in heap, stack, and static memory) and
Section~\ref{sec:lfcpa.arrays} adds the treatment of arrays and pointer
arithmetic to the formulation. Section~\ref{sec:unions} extends our
formulation to handle C style unions. Section~\ref{sec:conclusions}
concludes the paper.

\section{Original LFCPA}
\label{sec:original.lfcpa}

The definitions presented in this section have been excerpted from the original formulation~\cite{lfcpa}
and have been presented without any explanation.

Given relation $A\subseteq \pointer\times\var$ (either $\ain_n$ or
$\aout_n$) we first define an auxiliary extractor function
\begin{align}
  \must(A) & 
	= \bigcup_{x\in \pointer} \{x\} \times
  \left\{\begin{array}{c@{\hspace{1em}}l}
     \var & A\{x\} = \emptyset \vee A\{x\} = \{?\}           \\
     \{y\} & A\{x\} = \{ y \} \wedge y \neq \mbox{?}                \\
     \emptyset & \mbox{otherwise}
  \end{array}\right.
	\label{eq:must.new.def}
\end{align}

Extractor functions for statement $n$ are \text{$ \Def_n, \Kill_n,
\Ref_n \subseteq \pointer$} and \text{$\Pointee_n \subseteq \var $}. The
data flow values are \text{$\lin_n, \lout_n \subseteq \pointer$} and
\text{$\ain_n, \aout_n \subseteq \pointer \times \var$}.

The extractor functions are defined as follows. We assume that \text{$x,
y \in \pointer$} and \text{$a\in \var$}. $A$ abbreviates $\ain_n$.

\[
\begin{array}{|c|c|c|c|c|c|}
\hline
\multirow{2}{*}{Stmt.} 
	& \multirow{2}{*}{$\Def_n$}
	& \multirow{2}{*}{$\Kill_n$}
	& \multicolumn{2}{c|}{\Ref_n}
	& \multirow{2}{*}{$\Pointee_n$}
	\\ \cline{4-5}
	&
	&
	& \text{if } \Def_n \cap \lout_n \neq \emptyset\rule{0em}{1em}
	& \text{Otherwise}
	&
		\\ \hline\hline
\text{\em use\/} \; x
	& \emptyset
	& \emptyset
	& \rule[-.5em]{0em}{1.5em}
	 \{ x\}
	& \rule[-.5em]{0em}{1.5em}
	 \{ x\}
	& \emptyset
		\\ \hline
x = \& a
	& \{ x\}
	& \{ x\}
	& \rule[-.5em]{0em}{1.5em}
	 \emptyset
	& \rule[-.5em]{0em}{1.5em}
	 \emptyset
	& \{ a\}
		\\ \hline
x = y
	& \{ x\}
	& \{ x\}
	& \{y\}  
	& \emptyset
	& \rule[-.5em]{0em}{1.5em}
	 A\{y\} 
		\\ \hline
x = *y 
	& \{ x\}
	& \{ x\}
	& \{ y \} \cup (A\{y\} \cap \pointer)  \;\;
	& \emptyset
	& \rule[-.6em]{0em}{1.75em}
		A(A \{y\} \cap \pointer)
		\\ \hline
* x = y 
	& A\{x\} \cap \pointer \;
	 \rule[-.5em]{0em}{1.5em}
	& \must(A)\{x\} \cap \pointer 
	&  \{ x, y\}
	&  \{ x\}
	& A\{y\}
		\\ \hline
\text{other} \rule[-.25em]{0em}{1.25em}
	& \emptyset
	& \emptyset
	& \emptyset
	& \emptyset
	& \emptyset
		\\ \hline
\end{array}
\]

The data flow equations are:

\begin{eqnarray}
\lout_n & = & 
	\left\{
		\begin{array}{c@{\ \ \ \ }l}
			\emptyset 	& n \text{ is } \End{p}
				\\
			\displaystyle\bigcup_{s \in \Succ(n)} \lin_s
				& \text{otherwise}
		\end{array}
	\right.
	\label{eq:lout.exhaustive}
	\\
\lin_n & = & 
                \left(\lout_n - \Kill_n\right) \cup \Ref_n 
	\label{eq:lin.exhaustive}
	\\
\ain_n & = & 
	\left\{ \renewcommand{\arraystretch}{1.2}
		\begin{array}{c@{\ \ \ \ }l}
			\lin_n \!\times\! \{?\}
                                & n \text{ is } \Start{p}
				\\
		\left.
		\left(\displaystyle\bigcup_{p \in \pred(n)} \aout_p\right)
		\right|_{\mbox{$\lin_n$}}
				& \text{otherwise}
		\end{array}
	\right.
	\label{eq:ain.exhaustive}
	\\
\aout_n & = & 
\lrestrict{
                \left(\left(\ain_n  -  \left(\Kill_n \!\times\! \var\right)\right) \cup 
			\left( \Def_n \!\times\! \Pointee_n\right)\right)
}{\mbox{$\lout_n$}}
	\label{eq:aout.exhaustive}
\end{eqnarray}

\section{Extending LFCPA to Support Heap Memory and Structures}
\label{sec:lfcpa.heap}

\newcommand{\heap}{\text{\sf\em H\/}\xspace}
\newcommand{\fields}{\text{\sf\em F\/}\xspace}
\newcommand{\pfield}{\text{\sf\em pF\/}\xspace}
\newcommand{\npfield}{\text{\sf\em npF\/}\xspace}
\newcommand{\lRef}{\text{{\sf\em Lref}}\xspace} 
\newcommand{\rRef}{\text{{\sf\em Rref}}\xspace}
\newcommand{\hloc}{\text{{\sf\em heap}}\xspace} 
\newcommand{\allocsite}{\text{{\sf\em get\_heap\_loc}}\xspace} 
\newcommand{\lval}{\text{\sf\em lval\/}\xspace} 
\newcommand{\rval}{\text{\sf\em rval\/}\xspace} 
\newcommand{\addr}{\text{{\sf\em addr}}\xspace} 
\newcommand{\refp}{\text{\sf\em ref\/}\xspace} 
\newcommand{\derefp}{\text{\sf\em deref\/}\xspace} 
\newcommand{\lhs}{\text{\sf\em lhs\/}\xspace} 
\newcommand{\rhs}{\text{\sf\em rhs\/}\xspace} 
\newcommand{\closure}{\text{\sf\em closure\/}\xspace} 

We describe the proposed extensions by modelling pointer expressions
that involve structures and heap, and by defining the lattices and flow
functions. The data flow equations remain unchanged, only the flow
functions are redefined. 

\subsection{Modelling Pointer Expressions and the IR}

Original LFCPA has been formulated for pointers in stack and static
memory. Thus it has a simple model of memory and pointer expressions. 
The locations are named using variables and pointer expressions have only $*$ and $\&$
operators. In order to extend it to support heap data and structures on
stack, static area, and heap, we consider three kinds of pointers for a
C like language and introduce the sets that we use in our notation.

\begin{itemize}
\item A pointer variable \text{$x \in \pointer$} allocated on stack or
      in static data area.

\item A pointer contained in a heap location \text{$o_i \in \heap$}.

\item A pointer contained in a field $f$ of a struct variable. The      
      space for this variable may be allocated on stack, static area,   
      or heap.                                                          

      \begin{itemize}
      \item If accessed directly, such a pointer appears as \text{$x.f$} where
            \text{$x \in (\var-\pointer)$}, and \text{$f \in \pfield$}.
            In general, it can be a string like \text{$x.p.q.f$}
            where \text{$x \in (\var - \pointer)$}, $p$ and $q$ are
            non-pointer fields in \npfield, and \text{$f \in \pfield$}.
      \item When accessed through pointers, it appears as
            \text{$x\rightarrow f$} where \text{$x \in \pointer$},
            and \text{$f \in \pfield$}. It can be a sequence like
            \text{$x\rightarrow p\rightarrow q\rightarrow f$} where
            \text{$x \in \pointer$}, and \text{$p,q,f \in \pfield$}.
      \end{itemize}
      In general, we can have combinations of the two forms. For Java,
      however, we have only the latter form and \text{$\npfield =
      \emptyset$}.
\end{itemize}

We require the control flow graph form of the IR to be analysed.
Although we do not need the statements to be in a 3-address code format,
we assume the following two simplifications. In both the cases,
theoretically, it may be possible to formulate an analysis that does not
need the normalizations. However, it is not desirable for the reasons
described below.
\begin{itemize}
\item Expressions involving the `\&' operator.

      The operand of an {\em addressof\/} operator `\&' is required to
      be an l-value but its result is not an l-value. Hence expressions
      consisting of `\&' operators cannot be combined orthogonally.
      Multiple occurrences of `\&' are either illegal or are superfluous
      due to the presence of `$*$' and `$\rightarrow$' operators which
      would be required to make the expression semantically correct.
      We assume that the expressions involving `\&' are simplified so
      that there is a single `\&' which occurs in the beginning of an
      expression.\footnote{GCC simplifies pointer expressions containing
      `\&' in such a manner and we expect all compilers would do the same.}

      Intuitively, the `\&' and `$*$'/`$\rightarrow$' operators have
      ``opposite'' semantics in terms of the memory graph; the former
      identifies predecessors of a node in the memory graph while the
      latter identifies the descendants of a node in the memory graph.
      Combining both in the analysis is possible but would complicate
      the formulation.

\item Assignment statements $\alpha = \alpha'$ where both $\alpha$ and 
      $\alpha'$ are structures (their types must be same).

      We assume that such an assignment statement has been replaced by 
      the assignments generated by the closure as defined below. 
      The order of these statements is immaterial.
	\begin{align}
	\closure(\alpha = \alpha') &=
		\begin{cases}
			\displaystyle\bigcup \closure(\alpha.m = \alpha'.m)
			& \renewcommand{\arraystretch}{.8}%
				\begin{array}{@{}l}
				\alpha \text{ and $\alpha'$ are structures}
				\\
				\text{and $m$ is a field of } \alpha
				\end{array}
			\\
		\left\{\alpha = \alpha'\right\} & \text{otherwise} 
		\end{cases}
	\end{align}

      Formulating this as a part of analysis changes control flow graph
      and describing these changes would complicate the formulations.

      The other option is to handle this implicitly by describing the
      effect of the transformations. Our formulation analyses pointer
      expressions in terms of their parts. This amounts to dividing
      the expressions in smaller parts. Handling structure assignments
      requires us to create larger expressions from smaller expressions.
      While it is possible to combine both reductions and expansions in
      an analysis, it would complicate the formulation.
      
\end{itemize}

The following grammar\footnote{This grammar is ambiguous; we use the
precedences and associativities of the operators as defined for C to
disambiguate it.} defines a pointer expression $\alpha$ 
which may appear in other expressions, conditions,
or assignment statements in a program. These expressions are defined
in terms of \text{$x \in \var$}, \text{$f \in \pfield \cup \npfield$}, and
call to {\em malloc\/} function for memory allocation.
\begin{align}
\alpha & :=
	\text{\em malloc\/}
	\mid \&\beta
	\mid \beta
	\\
\beta & :=  x 
       \mid \beta.f 
       \mid \beta\rightarrow f 
       \mid *\beta  
\end{align}
If \text{$f\in \pfield$} or \text{$x \in \pointer$}, then $\beta$ is a
pointer. Besides, for \text{$ \beta\rightarrow f $} and \text{$ *\beta
$}, $\beta$ must be a pointer. In other cases, $\beta$ may not be a
pointer. When $\alpha$ is \text{\em malloc\/} or \text{$\&\beta$}, it
does not have an l-value can appear as a pointee only.

\subsection{Memory Model and the Lattice of Pointer-Pointee Relations}

Since the heap memory is unbounded, we abstract the heap locations based
on allocation sites. The memory chunk allocated by an
assignment statement \text{\em n:\; x = malloc$()$} is named $o_n$ where $n$ is the
label of the statement. We define \heap to contain such names. Note that $o_n$ is a compile time constant
and the set \heap fixed.

\newcommand{\structp}{\text{\sf\em S$_{p}$}\xspace}
\newcommand{\structm}{\text{\sf\em S$_{m}$}\xspace}
\newcommand{\genp}{\text{\sf\em G$_{p}$}\xspace}
\newcommand{\genm}{\text{\sf\em G$_{m}$}\xspace}
\newcommand{\rootp}{\text{\sf\em R\/}\xspace}
\newcommand{\source}{\text{\sf\em S\/}\xspace}
\newcommand{\target}{\text{\sf\em T\/}\xspace}
\newcommand{\inner}{\text{\sf\em inner\_level\/}\xspace}
\newcommand{\sibling}{\text{\sf\em same\_level\/}\xspace}
\newcommand{\coincides}{\text{\sf\em coincides\/}\xspace}

We define \structp to contain named locations corresponding to the pointers within structures, and \structm to
contain named locations for all members of structures.
For brevity, let \text{$ \rootp = (\var\!-\!\pointer)\! \cup\! \heap$} represent the root of a named location. 
\begin{align}
\structp & = \rootp \! \times\! \npfield^{*}\!\times \pfield
		\label{eq:struct.ptr}
			\\
\structm & = \rootp \! \times\! \npfield^{*} \!\times\! (\pfield\! \cup\! \npfield)
		\label{eq:struct.mem}
\end{align}
For convenience, we denote a named location consisting of a tuple \text{$(a,b,c,d)$} by concatenating the names
as \text{$a.b.c.d$}.
Some examples of named locations are: 
\begin{itemize}
\item \text{$a.g.h.f \in \structp$} where \text{$a \in \rootp$} (because \text{$a \in         
       \var-\pointer$}); \text{$g,h \in \npfield$}; and \text{$f \in \pfield$}.
\item  \text{$o_1.h.g \in \structm$} where \text{$o_1 \in \rootp$} 
      (because \text{$o_1 \in \heap$}), and \text{$g,h   \in \npfield$}. 
\end{itemize}
A field \text{$f \in \pfield$} cannot appear in the middle
because a pointer field cannot have subfields.                                                        

It is easy to see that a named locations consist of compile time constants.

\begin{figure}[t]
\begin{center}
\begin{tabular}{@{}c|c}
Code & Memory \\ \hline
\tt
\begin{tabular}{@{}c|c@{}}
\begin{tabular}{@{}l@{}}
typedef struct B
	\\
\{\hspace*{5mm}\ldots
	\\
\hspace*{5mm}struct B *f;
	\\
\} sB;
\\
typedef struct A 
	\\
\{\hspace*{5mm}\ldots
	\\
\hspace*{5mm}struct B g;
	\\
\} sA;
\end{tabular}
&
\begin{tabular}{@{}l@{}}
\phantom{1. }sA *a; \\
\phantom{1. }sB *x, *y, b; \\
1. a = (sA*) malloc \\
\hspace*{17mm}(sizeof(sA)); \\
2. y = \&a->g;  \\
3. b.f = y; \\
4. x = \&b; \\
5. return x->f->f;
\end{tabular}
\end{tabular}
&
\begin{tabular}{@{}c@{}}
\begin{pspicture}(0,2)(65,29)

\putnode{xn}{origin}{10}{4}{\psframebox{\rule{0mm}{3mm}\rule{6mm}{0mm}}}
\putnode{xn}{origin}{10}{4}{}
\putnode{w}{xn}{-7}{0}{$x$}
\putnode{an}{xn}{0}{11}{\psframebox{\rule{0mm}{3mm}\rule{6mm}{0mm}}}
\putnode{an}{xn}{0}{11}{}
\putnode{w}{an}{-7}{0}{$a$}
\putnode{yn}{an}{0}{11}{\psframebox{\rule{0mm}{3mm}\rule{6mm}{0mm}}}
\putnode{yn}{an}{0}{11}{}
\putnode{w}{yn}{-7}{0}{$y$}
\putnode{bn}{xn}{20}{4}{\psframebox{\begin{tabular}{c}
				$f$\\
				{\psframebox{\rule{3mm}{0mm}\rnode{bfn}{}\rule[-1.5mm]{0mm}{3mm}\rule{3mm}{0mm}}}
				\end{tabular}}}
\putnode{w}{bn}{-10}{6}{$b$}
\putnode{on}{yn}{44}{-14}{\psframebox{\begin{tabular}{c}
				$g$\\
				\rnode{ogn}{\psframebox{\begin{tabular}{c}
					$f$\\
				\rnode{ogfn}{\psframebox{\rule{0mm}{4mm}\rule{6mm}{0mm}}}
				\end{tabular}}}
					\\
				\end{tabular}}}
\putnode{w}{on}{-14}{9}{$o_1$}

\nccurve[angleA=0,angleB=190,arrowsize=2mm]{*->}{xn}{bn}
\nccurve[angleA=15,angleB=110,arrowsize=2mm,offsetB=-2,nodesepB=1]{*->}{yn}{ogn}
\nccurve[angleA=10,angleB=185,arrowsize=2mm]{*->}{bfn}{ogn}
\nccurve[angleA=30,angleB=160,arrowsize=2mm]{*->}{an}{on}

\end{pspicture}
\end{tabular}
\end{tabular}
\end{center}
\caption{Pointer expressions and their named locations.
The named location of $x$, \text{$x\rightarrow f$}, and
\text{$x\rightarrow f\rightarrow f$} are $x$, \text{$b.f$}, and 
\text{$o_1.g.f$} respectively, where $o_1$ is a heap location derived from its allocation site.
}
\label{fig:ptr.expr}
\end{figure}
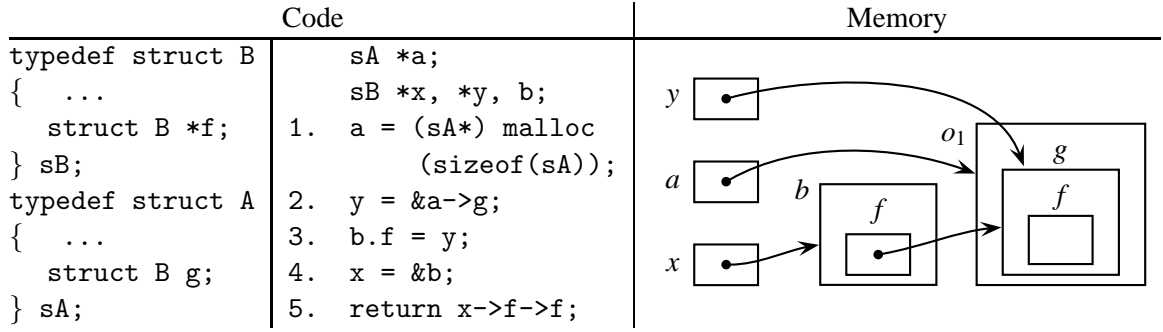

Let \source denote the set of all pointers and \target denote the set of all
pointees. Thus \source forms the source set and \target, the target set, in a
points-to relation \text{$A \subseteq \source\times T$}.
\begin{align}
\source & = \pointer 
		\; \cup \; \heap 
		\; \cup \; \structp
	\label{eq:ptr.src}
			\\
\target & = \var 	
		\; \cup \; \heap 
		\; \cup \; \structm
		\;  \cup \; \{?\}
	\label{eq:ptr.tgt}
\end{align}
Observe that \source does not contain pointer expressions but their named locations;
similarly, \target does not contain expressions describing pointees but
their named locations. Figure~\ref{fig:ptr.expr} illustrates the named locations of
pointer expressions.

The lattice for data flow variables $\ain_n/\aout_n$ is \text{$(2^{\source\times
\target},\; \supseteq)$} whereas the lattice for $\lin_n/\lout_n$ is
\text{$(2^\source,\; \supseteq)$}. The extractor functions $\Def_n, \Kill_n,
\Ref_n$ compute subsets of \source.

\subsection{Flow Functions}

Since our pointer expressions have a rich structure now, we define some auxiliary functions
including those that compute the l- and r-values of pointer expressions. These auxiliary
functions are then used to define the extractor functions which extract the pointers used
in a given statement in the IR.

\subsubsection*{Computing the l- and r-values of pointer expressions}

Given a points-to relation \text{$A \subseteq \source\times \target$} and an
expression $\alpha$ involving pointers, we define the following
auxiliary functions that evaluate a pointer expression $\alpha$ in the
environment of points-to relation $A$.
These functions are defined recursively using structural induction on the grammar 
rules thereby ensuring that they cover all possibilities.

We use the following notation: $\alpha$ and $\beta$ are pointers expressions;
field $f$ may be a pointer or non-pointer (i.e.
\text{$f \in \pfield \cup \npfield$})
and $\sigma$ is either a name in \var or \heap, or a string representation of the name tuple in \structp or \structm.

\begin{itemize}
\item {$\lval(\alpha,A)$} computes the set of possible l-values         
      corresponding to $\alpha$. 
Some pointer expressions such as $\&x$    
      and \text{\em malloc\/} do not have an l-value. 
\begin{align}
\lval(\alpha,A) & =  
		\begin{cases}
		\{ \sigma \} 	
				& (\alpha \equiv \sigma) \wedge (\sigma \in \var)
			\\
		\{ \sigma.f \mid \sigma \in \lval(\beta,A) \} 
				& \alpha \equiv \beta.f
			\\
		\left\{ \sigma.f \mid \sigma \in \rval(\beta,A), \sigma \neq ? \right\}
				& \alpha \equiv \beta\rightarrow f
			\\
		\left\{ \sigma \mid \sigma \in \rval(\beta,A), \sigma \neq ? \right\}
				& \alpha \equiv *\beta
			\\
		\emptyset	& \text{otherwise}
		\end{cases}
		\label{eq:lval.heap}
\end{align}
The pointer expressions \text{$\beta\rightarrow f$} and 
\text{$*\beta$} involve pointer indirection at the outermost level (i.e. pointees of
$\beta$ are read) and hence we use $\rval(\beta)$. This 
covers the case when $\beta$ is a pointer expression that does not have an l-value (eg.
\text{$(\&x)\rightarrow f$}, or \text{$*(\&x)$}).

\item {$\rval(\alpha,A)$} computes the set of possible r-values
      corresponding to $\alpha$. These are the contents (i.e. the pointees) of the l-values
      of $\alpha$. These are defined only if the l-values of $\alpha$ are pointers
      or $\alpha$ has no l-value but is $\&\beta$ or \text{\em malloc}.
\begin{align}
\rval(\alpha, A) & = 
		\begin{cases}
		\lval(\beta,A)
			& \alpha \equiv \&\beta
			\\
		\{ o_i \}	
			& \alpha \equiv \text{\em malloc} \wedge o_i = \allocsite()
			\\
		A(\lval(\alpha,A) \cap \source) 
			& \text{otherwise}
		\end{cases}
		\label{eq:rval.heap}
\end{align}
Function \text{$\allocsite()$} uses the label of the current statement to create a name 
for heap location allocated by statement. 
We leave the details implicit for simplicity.

\item \text{$\derefp(\alpha,A)$} computes the set of pointers that need
       to be read to reach the locations in $\lval(\alpha,A)$.
\begin{align}
\derefp(\alpha,A) & =  
		\begin{cases}
		\derefp(\beta,A) 
				& \alpha \equiv \beta.f
			\\
		\lval(\beta,A) \cup \derefp(\beta,A) 
				& (\alpha \equiv \beta\rightarrow f)  \vee (\alpha \equiv *\beta) 
			\\
		\emptyset	& \text{otherwise}
		\end{cases}
		\label{eq:derefp.heap}
\end{align}
Since pointer expressions \text{$\beta\rightarrow f$} and 
\text{$*\beta$} involve pointer indirection at the outermost level we
include the l-values of $\beta$ also in \derefp.

\item \text{$\refp(\alpha,A)$} computes the set of pointers which would
       be read to extract the values in $\rval(\alpha,A)$.
\begin{align}
\refp(\alpha,A) & =  
		\begin{cases}
		\derefp(\beta,A) 
			& \alpha \equiv \&\beta
			\\
		\derefp(\alpha,A) \cup \left(\lval(\alpha,A) \cap \source\right)
			\hspace*{10mm}
			&	\text{otherwise}
		\end{cases}
		\label{eq:refp.heap}
\end{align}
\end{itemize}

Figure~\ref{fig:aux.fn.exmp} illustrates these functions for 
some pointer expressions corresponding to the program fragment in Figure~\protect\ref{fig:ptr.expr}.

\begin{figure}[t]
\begin{center}
$
\begin{array}{|l|l|l|l|l|}
\hline
\multirow{2}{*}{$\alpha$}
	&
	\multicolumn{4}{c|}{
	A = \left\{
	\left(a, o_1 \right),
	\left(y, o_1.g \right),
	\left(x, b \right),
	\left(b.f, o_1.g \right)
	\left(o_1.g.f, ? \right)
	\right\}
	}
	\\ \cline{2-5}
	& \lval(\alpha,A)
	& \rval(\alpha,A)
	& \derefp(\alpha,A)
	& \refp(\alpha,A)
	\\ \hline\hline
x 
	& \{ x \}
	& \{ b \}
	& \emptyset
	& \{ x \}
	\\ \hline
a 
	& \{ a \}
	& \{ o_1 \}
	& \emptyset
	& \{ a \}
	\\ \hline
*a 
	& \{ o_1 \}
	& \emptyset
	& \{ a \}
	& \{ a, o_1 \}
	\\ \hline
*x 
	& \{ b \}
	& \emptyset
	& \{ x \}
	& \{ x, b \}
	\\ \hline
*y 
	& \{ o_1.g \}
	& \emptyset
	& \{ y \}
	& \{ y, o_1.g \}
	\\ \hline
b.f
	& \{ b.f \}
	& \{ o_1.g \}
	& \emptyset
	& \{ b.f \}
	\\ \hline
a \rightarrow g
	& \{ o_1.g \}
	& \emptyset
	& \{ a \}
	& \{ a, o_1.g \}
	\\ \hline
y \rightarrow f
	& \{ o_1.g.f \}
	& \{ ? \}
	& \{ y \}
	& \{ y, o_1.g.f \}
	\\ \hline
x \rightarrow f
	& \{ b.f \}
	& \{ o_1.g \}
	& \{ x \}
	& \{ x, b.f \}
	\\ \hline
x \rightarrow f \rightarrow f
	& \{ o_1.g.f \}
	& \{ ? \}
	& \{ x, b.f \}
	& \{ x, b.f, o_1.g.f \}
	\\ \hline
\end{array}
$
\end{center}
\caption{Examples of \lval, \rval, \derefp, and \refp, for some pointer expressions corresponding to Figure~\protect\ref{fig:ptr.expr}.}
\label{fig:aux.fn.exmp}
\end{figure}

For the original LFCPA, the auxiliary functions reduce to the following:
\begin{align}
\lval(\alpha,A) & =  
		\begin{cases}
		\{ x \} 	
				& (\alpha \equiv x) \wedge (x \in \var)
			\\
		\left\{ \sigma \mid \sigma \in \rval(x,A), \sigma \neq ? \right\}
				& \alpha \equiv *x
			\\
		\emptyset	& \text{otherwise}
		\end{cases}
		\\
\rval(\alpha, A) & = 
		\begin{cases}
		\lval(x,A) = \{ x \}
			& \alpha \equiv \&x
			\\
		A(\lval(\alpha,A) \cap \source) 
			& \text{otherwise ($\alpha\equiv *x$ or $\alpha\equiv x$)}
		\end{cases}
		\\
\derefp(\alpha,A) & =  
		\begin{cases}
		\lval(x,A) 
				& \alpha \equiv *x 
			\\
		\emptyset	& \text{otherwise}
		\end{cases}
		\\
\refp(\alpha,A) & =  
		\begin{cases}
		\derefp(x,A) 
			& \alpha \equiv \&x
			\\
		\derefp(\alpha,A) \cup \left(\lval(\alpha,A) \cap \source\right)
			&	\text{otherwise ($\alpha\equiv *x$ or $\alpha\equiv x$)}
		\end{cases}
\end{align}

\newcommand{\ma}{\text{\em mA\/}\xspace}

\subsubsection*{Extractor functions used in the data flow equations}
An abstract heap object may represent multiple concrete heap objects thus
prohibiting strong updates. The predicate \text{$\hloc(l)$} asserts
that a location $l$ is on heap.  
\begin{align}
\hloc(l) &
	\Leftrightarrow 
	   l \in \heap 
		\; \cup \; \heap\! \times\! \npfield^{*} \!\times\! (\pfield\! \cup\! \npfield)
\end{align}
Given a points-to relation \text{$A \subseteq \source\times \target$},
we define its must version using this predicate as follows:
\begin{align}
  \must(A) & 
	= \bigcup_{p\in \source} \{p\} \times
	\begin{cases}
     \target & \left(A\{p\} = \emptyset \vee A\{p\} = \{?\}\right)    \wedge \neg\hloc(p)      \\
     \{q\} & A\{p\} = \{ q \} \wedge q \neq \mbox{?} \wedge \neg\hloc(q)   \wedge \neg\hloc(p)                \\
     \emptyset & \mbox{otherwise}
  \end{cases}
	\label{eq:must.heap.def}
\end{align}

Let $A$ denote $\ain_n$ and \ma denote \text{$\must(A)$} . Then,
the extractor functions $\Def_n$, $\Kill_n$, $\Ref_n$, and $\Pointee_n$ are defined for various statements
as follows. 
\begin{itemize}
\item Pointer assignment statement \text{$\lhs_n = \rhs_n$}. We assume that this statement is type correct and both 
$\lhs_n$ and $\rhs_n$ are pointers.
\begin{align}
\Def_n & = \lval(\lhs_n,A) 
		\label{eq:def.heap} 
	\\
\Kill_n & =  
	\lval\left(\lhs_n, \ma\right)
		\label{eq:kill.heap}
	\\
\Ref_n & = 
		\begin{cases}
		\derefp(\lhs_n,A)   
			& \Def_n \cap \lout_n = \emptyset
			\\
		\derefp(\lhs_n,A) \cup \refp(\rhs_n,A)   \hspace*{4mm}
			& \text{otherwise}
		\end{cases}
	\\
\Pointee_n & =  
	\rval(\rhs_n, A)
\end{align}
Observe the use of  \ma (i.e. \text{$\must(A)$}) in the definition of $\Kill_n$.

\item {\em Use\/} $\alpha$ statement. This statement models all uses of pointers which are 
      not in a pointer assignment statement (eg. \text{\tt x->n = 10;} where {\tt n} is an integer field of a structure).
	\begin{align}
	\Def_n & = \Kill_n = \Pointee_n = \emptyset
		\\
	\Ref_n & = \refp(\alpha,A)
		\label{eq:ref.use.heap}
	\end{align}
\item Any other statement.
\begin{align}
	\Def_n  & = \Kill_n = \Ref_n = \Pointee_n = \emptyset
		\label{eq:anu.other.heap}
\end{align}
\end{itemize}

The data flow equations (\ref{eq:lout.exhaustive})$-$(\ref{eq:aout.exhaustive}) remain unchanged.

Observe that if we exclude heap and structures and restrict $\alpha$ to scalar pointers in \pointer, 
definitions \text{(\ref{eq:def.heap})$-$(\ref{eq:anu.other.heap})} of
$\Def_n$, $\Kill_n$, $\Ref_n$, and $\Pointee_n$ reduce to the definitions in the original LFCPA formulations.

Figure~\ref{fig:extractor.fn.exmp} shows the values of the extractor functions for the final round of analysis
(i.e., they are computed using the final liveness and points-to information).

\begin{figure}[t]
\begin{center}
$
\begin{array}{|c|l|l|l|l|}
\hline
\text{Stmt. $n$}
	& \Def_n
	& \Kill_n
	& \Ref_n
	& \Pointee_n
	\\ \hline\hline
1
	& \{ a \} 
	& \{ a \} 
	& \emptyset
	& \{ o_1 \}
	\\ \hline
2 
	& \{ y \} 
	& \{ y \} 
	& \{ a \} 
	& \{ o_1.g \} 
	\\ \hline
3 
	& \{ b.f \} 
	& \{ b.f \} 
	& \emptyset
	& \{ o_1.g \}
	\\ \hline
4 
	& \{ x \} 
	& \{ x \} 
	& \emptyset 
	& \{ b \}
	\\ \hline
5 
	& \emptyset 
	& \emptyset 
	& \{ x,b.f, o_1.g.f \} 
	& \emptyset
	\\ \hline
\end{array}
$
\end{center}

\caption{Extractor Functions for the statements in
Figure~\protect\ref{fig:ptr.expr}. These have been computed using the final
liveness and points-to information presented in 
Figure~\protect\ref{fig:data.flow.values}.}
\label{fig:extractor.fn.exmp}
\end{figure}

\section{Handling Pointer Arithmetic and Arrays}
\label{sec:lfcpa.arrays}

So far our pointers in aggregates are restricted to structures. Now we
extend them to include arrays as well as pointer arithmetic. Given an
arithmetic expression \text{$e \in \expr$}, the grammar for extended
pointer expressions is as follows. For simplicity we have considered only
$+$ operator for pointer arithmetic; $-$ operator can be handled by negating the
value of $e$.
\begin{align}
\alpha & :=
	\; \text{\em malloc\/}
	\;\mid\; \&\beta
	\;\mid\; \beta
       \;\mid\; \&\beta + e
	\\
\beta & :=  
       \; x 
       \;\mid\; \beta.f 
       \;\mid\; \beta\rightarrow f 
       \;\mid\; *\beta  
       \;\mid\; \beta[e]
       \;\mid\; \beta + e
\end{align}
In general, \text{$e \in \expr$} may be defined in terms of variables whose values may
not be known at compile time.
Let \text{$\eval{e} \in \const$} represent the evaluation of
expression $e$ where \const is a set of constants. If $e$ cannot be evaluated at compile time,  then
\text{$\eval{e} = \bot_{\eval{}}$} which
is included as a fictitious constant in \const. 

\begin{figure}[t]
\begin{center}
$
\begin{array}{|c|l|l|l|l|}
\hline
\text{Stmt. $n$}
	& \lin_n
	& \lout_n
	& \ain_n
	& \aout_n
	\\ \hline\hline
1 
	& \{ o_1.g.f \} 
	& \{ a,o_1.g.f \} 
	& \{ (o_1.g.f,?) \}
	& \{ (a,o_1),(o_1.g.f,?) \}
	\\ \hline
2 
	& \{ a,o_1.g.f \} 
	& \{ y,o_1.g.f \} 
	& \{ (a,o_1),(o_1.g.f,?) \} 
	& \{ (y, o_1.g),(o_1.g.f,?) \}
	\\ \hline
3 
	& \{ y,o_1.g.f \} 
	& \{ b.f,o_1.g.f \} 
	& \{ (y,o_1.g),(o_1.g.f,?) \} 
	& \{ (b.f,o_1.g),(o_1.g.f,?) \}
	\\ \hline
4 
	& \{ b.f,o_1.g.f \} 
	& \{ x,b.f,o_1.g.f \} 
	& \{ b.f,o_1.g),(o_1.g.f,?) \} 
	& 
		\renewcommand{\arraystretch}{.9}
		\begin{array}{@{}r@{}}
		\{ (x,b),(b.f,o_1.g),\strut
			\\
		(o_1.g.f,?) \}
		\end{array}
	\\ \hline
5 
	& \{ x,b.f,o_1.g.f \} 
	& \emptyset 
	& 
		\renewcommand{\arraystretch}{.9}
		\begin{array}{@{}r@{}}
		\{ (x,b),(b.f,o_1.g),\strut
			\\
		(o_1.g.f,?) \} 
		\end{array}
	& \emptyset
	\\ \hline
\end{array}
$
\end{center}

\caption{Final round of liveness and points-to analysis for the statements in
Figure~\protect\ref{fig:ptr.expr}.}
\label{fig:data.flow.values}
\end{figure}

\subsection{Memory Model and the Lattice of Pointer-Pointee Relations}
\label{sec:mem.model.array}

The inclusion of array accesses and pointer arithmetic goes
against the grain of our memory model which, so far, is not addressable.
In other words, we access a location by a name and not by an address
in the current model---functions \lval and \rval return a name rather than
an address. Such a model is based on an implicit assumption that two
different names refer to two different locations. The operators of
pointer expressions preserve this invariant. However, when we combine
names and pointer arithmetic or array indexing, the operators of
pointers expressions cannot preserve this invariant. As a consequence,
our model may not be able to find out, and hence cannot guarantee, that
the actual locations of pointers $x+4$ and $y$ are different. Such
checks would require an addressable model of memory where each name
(including the names created using \lval) is at an offset from a start
address; the offset for the location of a pointer expression can be
computed by reducing the pointer arithmetic and using the offsets of
fields and array elements.

While this may work well for static and stack locations, it would
not work for heap because we would not know the offsets of dynamic
memory allocation at compile time. Hence we discard the addressable
model\footnote{It may be possible to use separate memory models for
stack/static memory and heap.} and use a partially addressable model in
which only the locations within an array are addressable with respect to
the name of the array and no names overlap in memory even if they use
an offset. For ensuring soundness of may points-to analysis, we use the
following approximations:
\begin{itemize}
\item Whenever an offset for an array location cannot be computed at
      compile time, we view all accesses to the array as index-insensitive and treat the
      entire aggregate as a single variable. For soundness, read and
      write accesses would need different approximations based on this
      assumption.
	\begin{itemize}
	\item A read would be approximated as reading {\em any\/} location. 
	\item A write would be approximated as writing into {\em any\/}
      location for the purpose for generating liveness or points-to information and
      writing into {\em no\/} location for the purpose of 
      killing liveness or points-to information.
	\end{itemize}
\item When pointer arithmetic is used, we assume that the resulting location could coincide
      with 
	\begin{itemize}
	\item {\em any\/} location within the array if the pointer points to an array, or
	\item {\em any\/} of the named locations if the pointer does not hold the address of an array.
	\end{itemize}
\end{itemize}

With this model, we extend the structure pointers \structp and the structure member \structm
(equations (\ref{eq:struct.ptr}) and (\ref{eq:struct.mem})) to define
general pointers and members \genp and \genm. Since we can have
structures within arrays and arrays within structures, we allowing
field names to be interspersed with constant offsets.
Let $\const$ represent the set of constant offsets. Then,
\begin{align}
\genp & = \rootp \! \times\! 
	 \left(\const \cup \npfield\right)^{*}
	\times \left(\const \cup \pfield \right)
			\\
\genm & = \rootp \! \times\! 
	 \left(\const \cup \npfield\right)^{*}
	\times \left(\const \cup \pfield\cup\npfield\right)
\end{align}

\begin{figure}
\begin{center}
\begin{tabular}{c|c}
\begin{minipage}{50mm}
\begin{verbatim}
struct s1
{     int i;
      int *h;
};
struct s2
{    int y;
     struct s1 g[10];
};
\end{verbatim}
\end{minipage}
&
\begin{minipage}{50mm}
\begin{verbatim}
struct s3      
{    
     int x;
     struct s2 f;
};
struct s3 a[20][10];
\end{verbatim}
\end{minipage}
\end{tabular}
\end{center}
\caption{An example of complex nestings of arrays and structures. We model the location of pointer
\text{$a[7][3].f.g[5].h$} by the name
\text{$a.7.3.f.g.5.h$}. 
}
\label{fig:exmp.array.struct}
\end{figure}

Figure \ref{fig:exmp.array.struct} illustrates our modelling. Given two
names \text{$x.c_1.c_2$} and \text{$y.f.g$}, the computation of actual
memory locations for them would be different: $y$ is a structure and a
compiler would simply add the offsets of $f$ and $g$ to the address of
$y$ to get the actual location of \text{$y.f.g$}. However, in the name
\text{$x.c_1.c_2$}, $x$ is an array (because it is followed by a number
$c_1$) and the array address calculation performed by the compiler is
not just addition of $c_1$ and $c_2$. Instead, $c_1$ will have to be
multiplied by the number of columns and then $c_2$ will be added.

The offsets appearing in the named locations  are unscaled; we do not use a scaling factor based on 
the size of the data type. Our goal is to uniquely identify locations for pointer-pointee relations
rather than for accessing the memory. For the latter, the offsets will have to be scaled up using data
type. Since we assume a typed IR, this information should be available in the preceding field/variable name
in the list.

Observe that the named locations continue to be compile time constants.

The definitions of source and target sets remain same as equations
(\ref{eq:ptr.src}) and (\ref{eq:ptr.tgt}) except that \structp and \structm
are replaced by \genp and \genm respectively.
\begin{align}
\source & = \pointer 
		\; \cup \; \heap 
		\; \cup \; \genp
			\\
\target & = \var 	
		\; \cup \; \heap 
		\; \cup \; \genm
		\;  \cup \; \{?\}
\end{align}

\subsection{Extractor Functions}

The revised definition of \lval appears below. When compared with
equation (\ref{eq:lval.heap}), it is clear that the only change is
handling an array access; the result of pointer arithmetic does not have
an l-value. The l-value of a pointer
expression \text{$\beta[e]$}, is defined by appending the evaluation of
$e$ to the l-value of $\beta$. In case of multi-dimensional arrays, the
l-value of $\beta$ would already have a suffix containing some offsets.
The new case in the definition of \lval has been marked in blue.
\begin{align}
\lval(\alpha,A) & =  
		\begin{cases}
		\{ \sigma \} 	
				& (\alpha \equiv \sigma) \wedge (\sigma \in \var)
			\\
		\{ \sigma.f \mid \sigma \in \lval(\beta,A) \} 
				& \alpha \equiv \beta.f
			\\
		\left\{ \sigma.f \mid \sigma \in \rval(\beta,A), \sigma \neq ? \right\}
				& \alpha \equiv \beta\rightarrow f
			\\
		\left\{ \sigma \mid \sigma \in \rval(\beta,A), \sigma \neq ? \right\}
				& \alpha \equiv *\beta
			\\
			\blue
		\{ \sigma.\eval{e} \mid \sigma \in \lval(\beta,A) \} 
				& 
			\blue
				\alpha \equiv \beta[e]
			\\
		\emptyset	& \text{otherwise}
		\end{cases}
\end{align}
We allow $\bot_\eval{}$ to appear in a name string. In such a situation, it 
is interpreted as {\em any\/} location in that dimension of the array (as described
in Section~\ref{sec:mem.model.array}) except for \must and \Kill where
it is interpreted as {\em no\/} location as defined in equations
(\ref{eq:must.array.def}), and (\ref{eq:kill.array}) below.

The default case of \rval covers the array accesses and we only need
to add rules for handling pointer arithmetic. The most common use of
pointer arithmetic is to access array elements through pointers. In
such cases, pointer increments are well defined and pointer arithmetic
has a predictable behaviour. In other cases of pointer arithmetic, it is
difficult to find out the exact r-values of a pointer expression.

Consider a pointer expression \text{$x + c$}. If $x$ points to an array
location, the name of its r-value would have constant offset as its suffix 
(because the r-value of $x$ is an array location). In order to discover
the r-value of \text{$x + c$}, we simply need to add the scaled value of
$c$ to the offset with a scaling factor governed by the size of the type
of values held by the array. This is easily generalized to \text{$\beta
+ e$} in the definition of \rval. 
In all other cases of \text{$\beta + e$}, we
approximate the r-values by {\em any\/} location denoted by the universal
set of pointees \target. Similarly, the r-values of \text{$\&\beta + e$} are 
also approximated by the universal set of pointees \target.
The new cases have been marked in blue.
\begin{align}
\rval(\alpha, A) & = 
		\begin{cases}
		\lval(\beta,A)
			& \alpha \equiv \&\beta
			\\
		\{ o_i \}	
			& \alpha \equiv \text{\em malloc} \wedge o_i = \allocsite()
			\\
		\blue
		\target
				& 
		\blue
				\left(\alpha \equiv \beta + e \right) \wedge
				\left(\exists \,\sigma \in \rval(\beta, A), 
				\sigma \not\equiv \sigma'.c, \sigma' \in \target, c \in \const\right)
			\\
		\blue
			\displaystyle 
			\bigcup \; \{ \sigma.(c + \eval{e}) \} \rule[-.85em]{0em}{2.1em}
				& 
		\blue
				(\alpha \equiv \beta + e) 
				\wedge
				(\sigma.c \in \rval(\beta,A))
				\wedge
				(c \in \const)
			\\
		A(\lval(\alpha,A) \cap \source) 
			& \text{otherwise}
		\end{cases}
\end{align}

Since the default value of \derefp is $\emptyset$, we need to add the cases 
\text{$\beta[e]$} and \text{$\beta + e$} explicitly.
\begin{align}
\derefp(\alpha,A) & =  
		\begin{cases}
		\derefp(\beta,A) 
				& \left(\alpha \equiv \beta.n\right)
				\blue
					\vee
				 \left(\alpha \equiv \beta[e]\right)
					\vee
				 \left(\alpha \equiv \beta + e\right)
			\\
		\lval(\beta,A) \cup \derefp(\beta,A) 
				& (\alpha \equiv \beta\rightarrow m)  \vee (\alpha \equiv *\beta) 
			\\
		\emptyset	& \text{otherwise}
		\end{cases}
\end{align}
The pointers read to reach the pointees of \text{$\beta + e$} are included in the definition of
\rval as shown below.
\begin{align}
\refp(\alpha,A) & =  
		\begin{cases}
		\derefp(\beta,A) 
			& (\alpha \equiv \&\beta) 
				\blue
			\vee
			(\alpha \equiv \&\beta + e)
			\\
		\blue
		\refp(\beta,A) 
			& 
			\blue
			\alpha \equiv \beta + e
			\\
		\derefp(\alpha,A) \cup \left(\lval(\alpha,A) \cap \source\right)
			&	\text{otherwise}
		\end{cases}
\end{align}

The definition of \Kill should exclude the points-to relations of
approximated arrays. We achieve this by defining a by a predicate
$\app(l)$ which asserts that $l$ is a heap location or involves
$\bot_\eval{}$. The revised definitions of \Kill and \must are:
\begin{align}
  \must(A) & 
	= \bigcup_{p\in \source} \{p\} \times
	\begin{cases}
     \target & \left(A\{p\} = \emptyset \vee A\{p\} = \{?\}\right)    \wedge \blue \neg\app(p)      \\
     \{q\} & A\{p\} = \{ q \} \wedge q \neq \mbox{?} \wedge \blue \neg\app(q)   \wedge \neg\app(p)                \\
     \emptyset & \mbox{otherwise}
  \end{cases}
	\label{eq:must.array.def}
	\\
\Kill_n & =  
	\{ \sigma \mid \sigma \in \lval\left(\lhs_n, \ma\right), \neg \app(\sigma) \}
		\label{eq:kill.array}
\end{align}

All other definitions remain same except that
\text{$\bot_\eval{}$} is interpreted as {\em any\/} location.

Figure~\ref{fig:exmp.pointer.arith} illustrate compile time names for pointer expressions involving arrays and pointer arithmetic.

\begin{figure}[t]
\begin{center}
\begin{tabular}{l|l}
\begin{minipage}{40mm}
\begin{verbatim}
char *** t, ** s; 
char * q[10], p;

t = &s;
s = &q[3];
q[8] = &p;

\end{verbatim}
\end{minipage}
&
\begin{minipage}{110mm}
\raggedright
Pointer $s$ points to an array. The points-to information after these statements is
\text{$\{(t, s), (s, q.3), (q.8, p) \}$}. 
\begin{itemize}
\item The
\lval of pointer expressions 
\text{$*(*t + 5)$},
\text{$*(s + 5)$},
\text{$*(\&q[3] + 5)$}, and
\text{$q[8]$} after these statements is
$\{q.8\}$. 
\item The \rval of the same expressions is $\{ p\}$.
\end{itemize}
\end{minipage}
\\
\\ \hline
\\
\begin{minipage}{40mm}
\begin{verbatim}
int *** a, ** b;
int * c, * d;

a = &b;
b = &c;
\end{verbatim}
\end{minipage}
&
\begin{minipage}{103mm}
\raggedright
The points-to information after these statements is
\text{$\{\tt (a, b), (b, c) \}$}.
\begin{itemize}
\item The \lval of pointer expressions 
	\text{$*(*a + 5)$},
	\text{$*(b + 5)$}, and
	\text{$*(\&c + 5)$} is 
	$\target - \{?\}$. 
\item The \rval of the same expressions is  \target.
\end{itemize}
\end{minipage}
\end{tabular}
\end{center}
\caption{Examples of evaluations of pointer expressions involving arrays and pointer arithmetic.}
\label{fig:exmp.pointer.arith}
\end{figure}

\section{Handling Unions}
\label{sec:unions}

With support for handling structures, handling C style unions becomes straight forward.
We conservatively assume that when $\beta.f$ refers to access of field $f$ in
a union, it could coincide with 
\begin{itemize}
\item {\em any\/} field name for the purpose of generating the liveness and points-to information, and with
\item {\em no\/} field name for the purpose of killing the information.
\end{itemize}

The first requirement is served by approximating a union field-insensitively---we drop the field names
appearing in a union. 
The second requirement is served by ensuring that \text{$\isunion(\sigma) \Rightarrow \app(\sigma)$}
for equations (\ref{eq:must.array.def}) and~(\ref{eq:kill.array}).

The revised definition of \lval uses a
predicate $\isunion(\sigma)$ which asserts that $\sigma$ is a union. The 
changes have been marked in blue.
\begin{align}
\lval(\alpha,A) & =  
		\begin{cases}
		\{ \sigma \} 	
				& (\alpha \equiv \sigma) \wedge (\sigma \in \var)
			\\
		\renewcommand{\arraystretch}{.9}\begin{array}{@{}r@{}}
		\{ \sigma.f \mid \sigma \in \lval(\beta,A) {\blue, \neg\isunion(\sigma)} \} 
			\\
			\blue
			\cup\;
		\{ \sigma \mid \sigma \in \lval(\beta,A), \isunion(\sigma) \} 
		\end{array}
				& \alpha \equiv \beta.f 
				\rule[-1.25em]{0em}{3em}
			\\
		\renewcommand{\arraystretch}{.9}\begin{array}{@{}r@{}}
		\left\{ \sigma.f \mid \sigma \in \rval(\beta,A), \sigma \neq ? {\blue, \neg\isunion(\sigma)} \right\}
			\\
			\blue
			\cup\;
		\left\{ \sigma \mid \sigma \in \rval(\beta,A), \sigma \neq ?, \isunion(\sigma) \right\}
		\end{array}
				& \alpha \equiv \beta\rightarrow f
				\rule[-1.em]{0em}{2em}
			\\
		\left\{ \sigma \mid \sigma \in \rval(\beta,A), \sigma \neq ? \right\}
				& \alpha \equiv *\beta
			\\
		\renewcommand{\arraystretch}{.9}\begin{array}{@{}r@{}}
		\{ \sigma.\eval{e} \mid \sigma \in \lval(\beta,A) {\blue, \neg\isunion(\sigma)} \} 
			\\
			\blue
			\cup\;
		\{ \sigma \mid \sigma \in \lval(\beta,A), \isunion(\sigma) \} 
			\end{array}
				& 
				\alpha \equiv \beta[e]
				\rule[-.75em]{0em}{2.25em}
			\\
		\emptyset	& \text{otherwise}
		\end{cases}
\end{align}

Figure~\ref{fig:exmp.union} illustrates approximate names for unions.

\begin{figure}[t]
\begin{center}
\begin{tabular}{c|c}
\begin{minipage}{40mm}
\begin{verbatim}

union {
    struct {
        int * g;
    } f[30];
    int * h[20];
} a, b, * c[10];

c[4] = &a;
c[5] = &b;

\end{verbatim}
\end{minipage}
&
\begin{minipage}{105mm}
\begin{itemize}
\item The \lval of the following pointer expressions evaluates to {\tt a}: 

	{\tt a.f[0].g}, {\tt a.h[1]}, {\tt c[4]->f[2].g}, and {\tt c[4]->h[3]}. 

\item The \lval of the following pointer expressions evaluates to {\tt b}: 

{\tt b.f[6].g}, {\tt b.h[7]}, {\tt c[5]->f[8].g}, and {\tt c[5]->h[9]}.
\end{itemize}
\end{minipage}
\end{tabular}
\end{center}
\caption{An example of complex nestings of arrays, structures, and unions to
illustrate accesses of union fields and their compile time names created using field-insensitive approximation.}
\label{fig:exmp.union}
\end{figure}

\section{Conclusions}
\label{sec:conclusions}

With a suitable choice of naming conventions in terms of compile time
constants, and a suitable choice of functions to compute their l- and
r-values, extending LFCPA to support heap memory, structures, arrays,
and pointer arithmetic seems a relatively a straight forward extension.

We have chosen to retain the spirit of declarative formulation of the
original LFCPA and have not addressed the issue of efficient algorithms
for the formulations. 
We have described our
extensions in the intraprocedural setting. It remains to be seen
whether it is feasible to extend them to interprocedural
level using the default method of value contexts~\cite{%
Khedker.UP.Karkare.B:2008:Efficiency-Precision-Simplicity,%
Khedker.UP.Sanyal.A.Karkare.B:2009:Data-Flow-Analysis,%
Padhye:2013:IDF:2487568.2487569}
as was done in the original LFCPA 
or whether some additional issues need to be addressed. Further,
implementation of this method and empirical measurements would be a
non-trivial exercise and is left as future work.

We have handled heap memory using the allocation site
based abstraction in this paper. It would be interesting
to see how the use-site based abstraction as defined in
HRA~\cite{Khedker.UP.Sanyal.A.Karkare.A:2007:Heap-reference-analysis}
can be used in our extensions. Besides, the possibility of 
using an addressable model for stack and static memory with some
other model for heap can also be explored.

\section*{Acknowledgments}

We thank Swati Jaiswal and Pritam Gharat for their feedback in the early stages of the formulations in this paper.

\bibliography{general-lfcpa}

\end{document}